# Can Galactic chemical evolution explain the oxygen isotopic variations in the Solar System?

Maria Lugaro[1], Kurt Liffman[2], Trevor R. Ireland[3], Sarah T. Maddison[4]

[1]Monash Centre for Astrophysics (MoCA), Building 28, Monash University, Clayton, VIC 3800, Australia. Email: maria.lugaro@monash.edu

[2]CSIRO/MSE, P.O. Box 56, Highett, Victoria 3190, Australia

[3]Planetary Science Institute and Research School of Earth Sciences, The Australian National University, Canberra, ACT 0200, Australia

[4]Centre for Astrophysics & Supercomputing, Swinburne University, H39, PO Box 218, Hawthorn, VIC 3122, Australia

**ABSTRACT**

A number of objects in primitive meteorites have oxygen isotopic compositions that place them on a distinct, mass-independent fractionation line with a slope of one on a three-isotope plot. The most popular model for describing how this fractionation arose assumes that CO self-shielding produced $^{16}$O-rich CO and $^{16}$O-poor $H_2O$, where the $H_2O$ subsequently combined with interstellar dust to form relatively $^{16}$O-poor solids within the Solar Nebula. Another model for creating the different reservoirs of $^{16}$O-rich gas and $^{16}$O-poor solids suggests that these reservoirs were produced by Galactic chemical evolution (GCE) if the Solar System dust component was somewhat younger than the gas component and both components were lying on the line of slope one in the O three-isotope plot. We argue that GCE is not the cause of mass-independent fractionation of the oxygen isotopes in the Solar System. The GCE scenario is in contradiction with observations of the $^{18}O/^{17}O$ ratios in nearby molecular clouds and young stellar objects. It is very unlikely for GCE to produce a line of slope one when considering the effect of incomplete mixing of stellar ejecta in the interstellar medium. Furthermore, the assumption that the Solar System dust was younger than the gas requires unusual timescales or the existence of an important stardust component that is not theoretically expected to occur nor has been identified to date.

*Subject Headings:* Galaxy: abundances - Solar System: formation, oxygen isotopes

## 1. Introduction

Oxygen is produced by stellar nucleosynthesis and is the third most abundant element in the Solar System after hydrogen and helium, which were produced during the Big Bang. The chemical abundance of O in the precursor molecular cloud (MC) of the Sun was built up by the generations of stars that predated the birth of the Sun. This contribution can be calculated using Galactic chemical evolution (GCE) models, which can also be tested against spectroscopic observations of O in interstellar clouds and stars of different metallicities. The solar O abundance agrees with that observed in the solar neighborhood (see Sec. 4.2 of Asplund et al. 2009 and references therein), particularly so when allowing the Sun to have migrated from an original birth location 2-3 kpc closer to the Galactic centre (Nieva & Przybilla 2012). Most of the O in the Universe is in the form of $^{16}O$ (99.8% in the Solar System), an extremely stable isotope with double magic number of both protons and neutrons (Z=N=8). This nucleus is produced by α captures during He and Ne burning in massive ( >10 $M_\odot$) stars. These stars end their lives as core-collapse supernovae (SNII) and are responsible for the production of the "α elements", i.e., intermediate-mass elements comprised mostly by nuclei with an integer number of α particles, e.g., $^{16}$O, $^{20}$Ne, $^{24}$Mg, and $^{28}$Si. These nuclei are produced directly from the initial H and He present in the star and are independent on the initial stellar metallicity, i.e., they are a result of "primary" nucleosynthesis.

The minor isotopes of O, $^{17}$O and $^{18}$O, have much lower abundances than $^{16}$O, representing 0.04% and 0.2% of Solar System O, respectively. As outlined in detail by Meyer et al. (2008), $^{18}$O is predominantly a product of He burning in massive stars via the $^{14}$N$(α,γ)^{18}$F$(β^+)^{18}$O reaction chain, while $^{17}$O is a product of H burning via the $^{16}$O$(p,γ)^{17}$F$(β^+)^{17}$O reaction chain in low- and intermediate-mass stars (<10 M$_\odot$) as well as nova outbursts due to the accretion of material onto a white dwarf from a stellar companion in a binary system. These nuclei cannot be produced directly from the initial H and He present in the star and their production depends on the stellar metallicity, i.e., they are a result of "secondary" nucleosynthesis. The $^{17}$O/$^{16}$O and $^{18}$O/$^{16}$O ratios can be observed around cool red giant stars and in MCs and young stellar objects (YSOs) using vibration-rotation bands of the different isotopologues of the CO molecules. Oxygen is also a major constituent of dust, where a large number of meteoritic stardust grains are oxides and silicates (e.g., Nittler et al. 1997, Mostefaoui & Hoppe 2004, Nguyen et al 2010). These are analyzed for their O isotopic compositions via mass spectrometry to high precision and show large isotopic anomalies up to orders of magnitude with respect to the bulk of the Solar System. These data can be used to study stellar nucleosynthesis and GCE (e.g., Nittler 2009).

In the Solar System, the relative abundances of the stable oxygen isotopes have been measured to high precision in a large number of different materials (see Fig. 1, Fig. 4 of McKeegan et al. 2011, and Ireland 2012 for a review). The O isotopic ratios are usually expressed as permil deviations $\delta^{17}O$ and $\delta^{18}O$ from standard mean ocean water (SMOW):

$$\delta^i O = 10^3 \left( \frac{\left(^i O/^{16}O\right)}{\left(^i O/^{16}O\right)_{SMOW}} - 1 \right), \qquad (1)$$

where $i$ = 17 or 18. On the plot of $\delta^{17}O$ against $\delta^{18}O$ (Fig. 1) the Sun is enriched in $^{16}O$ by ~ 6% as measured in the solar wind collected by the Genesis mission[1] (McKeegan et al. 2011). All samples from the Earth, the Moon, Mars, and asteroid parent bodies lie very close, within 0.5%, to the line passing through the zero point (i.e., terrestrial) with a slope of approximately 0.52, the terrestrial fractionation line (TFL). This line arises due to equilibrium and kinetic processes that depend on the difference in mass between the oxygen isotopes (Young et al. 2002). In contrast, the major components of primitive chondrites - such as calcium–aluminum rich inclusions (CAIs), chondrules, amoeboid olivine aggregates (AOAs), and fine-grained matrices - plot along the slope-1 line (S1L) that connects the Earth and the Sun compositions (Fig. 1; Clayton et al. 1973; Young & Russell 1998). CAIs and AOAs are more $^{16}O$-rich than fine-grained matrices and chondrules (Yurimoto et al. 2008). The unity slope of this line strongly characterizes it as representing the compositions obtained by mixing between an $^{16}O$-rich reservoir and an $^{16}O$-poor reservoir. The $^{16}O$-rich reservoir is represented by the CAIs and the Sun, probably recording the composition of the gas component of the Solar Nebula. The $^{16}$O-poor reservoir is represented by the planets, probably recording the composition of the dust component of the Solar Nebula. The formation mechanism of these two reservoirs remains controversial.

---

[1] It should be noted that the measured solar wind O isotopic composition lies to the left of the S1L. The current favored, but not proven, interpretation is that the true solar composition lies on the S1L and the measured isotopic ratios were shifted due to fractionation processes during ionization and acceleration of the solar wind.

The dust component of the Solar Nebula was composed of dust formed in different environments. For sake of clarity we define here the terminology that will be used throughout the paper when referring to dust of different origins present in the Solar System. We will call *stardust* the meteoritic dust mentioned above that shows large isotopic anomalies up to orders of magnitude with respect to the bulk of the Solar System material. Stardust is believed to have formed in stellar envelopes and ejecta and to have preserved its original composition until today. As such, it carries the signature of nucleosynthesis, mixing, and condensation in stars (Clayton & Nittler 2004). Stardust of size from 20 μm down to ~ 100 nm has been discovered in the matrix of primitive meteorites. Stardust has low abundances, of the order of several to several hundred ppm. It is not yet known if stardust of smaller size is more abundant. Based on detailed models of the evolution of dust in the Galaxy, Zhukovska et al. (2008) have shown that the vast majority of the dust present at the time of the formation of the Solar System was likely not stardust, but had formed in MCs[2]. We call this *MC dust*. Some of the stardust and MC dust that were present at the time and place of Sun formation could have been destroyed during the formation of our Solar System by vaporization, sputtering, erosion, etc. At the same time new dust was forming in the Solar System directly from the gas phase. This dust we refer to as *Solar System condensates*. For sake of completeness we note that several processes could have further modified the dust present in the Solar System both preserving it as a closed system - in the case of melting, annealing, etc - and potentially involving interaction with gas and other dust - in the case of fragmentation, exchange, etc.

---

[2] In general, that most of the interstellar dust is not stardust but it is formed in MC is indicated by observational and theoretical studies that show that the timescale for replenishing interstellar dust is longer than that for destroying it (e.g., Gehrz 1989).

**1.2 Hypotheses for the S1L**

The first interpretation following the discovery of $^{16}$O-rich minerals in meteorites was that these materials contained the direct signature of nucleosynthetic events that rendered them "exotic", i.e., different from the bulk of the Solar Nebula (Clayton et al. 1973). Such an explanation could hold if the compositions of the Sun and Earth were close, instead, the Genesis mission revealed that the Sun is ~6% more $^{16}$O-rich than the Earth. While the meteoritic materials discovered by Clayton et al. (1973) lie on a line of slope ~0.95, which indicate an anomaly to be present also in the $^{17}$O/$^{18}$O ratio requiring a nuclear effect, Young & Russell (1998) have demonstrated that the data are in agreement with material originally lying on the S1L, and that the observed deviations are due to kinetic mass-dependent fractionation effects enriching the heavy O isotopes along lines of slope ~0.5. This indicates that the variations in meteoritic materials can be produced primarily as variations in $^{16}$O, with an offset from slope 1 being produced by kinetic fractionation and mixing.

The alternative hypothesis is that the O isotopic variations in the Solar System have a chemical rather than nuclear origin. The current most popular model involves "CO self-shielding" (see Ireland 2012 for a review). This model uses the property that ultraviolet (UV) radiation at specific wavelengths has sufficient energy to break carbon monoxide molecules, where the three oxygen isotopologues of CO ($C^{16}O$, $C^{17}O$ and $C^{18}O$) require slightly different photon energies to break the C-O bond. During the passage of light through MCs or YSO disks, the photons that can dissociate $C^{16}O$ are attenuated well before those required to dissociate $C^{17}O$ and $C^{18}O$ (hence the name self-shielding). Photodissociation and self-shielding (PSS) increases the number of $^{17}O$ and $^{18}O$ radicals in the cores of MCs and/or close to the surface of YSO disks. These can react with H to form $^{16}O$-poor water, which then reacts with refractory materials to produce the $^{16}O$-poor dust component, while CO, i.e., the gas component, becomes sympathetically enriched in $C^{16}O$. Different scenarios for PSS have been developed by Clayton (2002), Yurimoto & Kuramoto (2004), and Lyons & Young (2005). While PSS offers a compelling mechanism to explain $^{16}O$ variability, all scenarios have potential problems, related in particular to the exact mechanisms and timescales by which the $^{16}O$-poor water reacts with refractory materials to produce the $^{16}O$-poor dust. Several of these difficulties are discussed in detail by Krot et al. (2010).

Another model for the formation of the $^{16}O$-rich and $^{16}O$-poor reservoirs leading to the existence of the S1L comes from GCE (e.g., Jacobsen et al. 2007, Meyer 2009, Krot et al. 2010). In a very basic model the abundance of the primary $^{16}O$ increases linearly with time (Clayton 1988), while the abundances of the secondary $^{17}O$ and $^{18}O$ increase quadratically with time (Clayton & Pantelaki 1986). As a consequence, the galactic abundance ratios of $^{17}O/^{16}O$ and $^{18}O/^{16}O$ are expected to increase linearly with time. GCE can provide an explanation for the S1L if (1) the dust component of the Solar System derives directly from stardust or MC dust that was somewhat younger, hence had $^{17}O/^{16}O$ and $^{18}O/^{16}O$ ratios higher, than the gas component, and (2) the GCE evolution of the $^{18}O/^{17}O$ ratio is constant, matching the slope of the S1L. The aim of this paper is to analyze in detail these two requirements in relation to current GCE models and observations to determine if the S1L can be explained using GCE. We note that some of the points considered below are also discussed by Nittler & Gaidos (2012).

## 2. Models and observation of the $^{18}O/^{17}O$ ratio

The simple GCE considerations outlined above for the O isotopes indicate that, in principle, the assumption that the GCE evolution of the $^{18}O/^{17}O$ ratio is constant may be correct. In fact, early GCE models (e.g., Timmes et al. 1995) suggested that it is possible that the $^{18}O/^{16}O$ and $^{17}O/^{16}O$ ratios evolve at an equal pace and hence $^{18}O/^{17}O$ stays constant (see their Fig. 12). However, the situation is more complex. As briefly outlined in Sec. 1, even though both $^{17}O$ and $^{18}O$ are secondary isotopes, they are produced by entirely different processes, acting on different seed nuclei, and occurring in different stellar sources. A most important difference between these production sites is the timescale at which they start to contribute to the Galactic abundances. The massive stars that produce $^{18}O$ have much shorter lifetimes (~ Myr) than the low- and intermediate-mass stars and novae that produce $^{17}O$ (~ Gyr). Thus, when all possible stellar sources of the O isotopes are included in GCE models and their lifetimes are properly taken into account, there is no reason to expect *a priori* that $^{18}O/^{17}O$ evolves as a constant.

As an example, Figure 2 shows the most recent calculations of the GCE evolution of the O isotopic ratios (Kobayashi et al. 2011). The evolution of the O ratios on the O three-isotope plot is represented as a function of [Fe/H]=$\log_{10}\{$(Fe/H)/(Fe/H)$_\odot\}$, which can be used as a proxy for time since [Fe/H] increases with time as in Fig. 11b of Kobayashi et al. (2011). By definition [Fe/H] = 0 when the Solar System formed. For comparison, we also plot the S1L and a line of slope ½ representing the TFL. The slope of the GCE line from Kobayashi et al. (2011) suggests that early in Galactic history production of $^{18}$O was favored with respect to production of $^{17}$O, but as time progressed $^{17}$O production exceeded $^{18}$O production, resulting in a line with the approximate slope of 1.56. The reason for this is that the model of Kobayashi et al. (2011) includes the contribution of low- and intermediate-mass stars, which were not included in the model of Timmes et al. (1995). These stars produce $^{17}$O by H-burning while on the main sequence and carry it to the stellar surface in the red giant phase via the first dredge-up. Furthermore, if the initial stellar mass is greater than ~ 4 $M_\odot$ (the exact value depending on the metallicity), proton captures at the base of the convective envelope during the asymptotic giant branch (AGB) phase (i.e., "hot bottom burning") further increase the yield of $^{17}$O from these stars. As discussed in detail by Kobayashi et al. (2011), AGB stars in this mass range contribute significantly to the $^{17}$O abundance in the Galaxy. The result shown in Figure 2 indicates that this GCE model does not match the solar $^{17}$O/$^{16}$O ratio since the predicted value at solar metallicity is ~40% higher than observed. If the contribution of novae were also included in the models we would expect the $^{17}$O/$^{16}$O ratio to further increase relative to the $^{18}$O/$^{16}$O ratio (see also discussion in Romano & Matteucci 2003). GCE models have many uncertainties, e.g., stellar yields, the star formation rate, the initial mass function, etc. Traditionally, a GCE model is considered successful when it matches the solar isotopic abundances within a factor of two (e.g., Fig. 3 of Timmes et al. 1995). In our context, this means that the GCE line shown in Figure 2 should not be considered as a final accurate solution, but rather as an illustrative example that we should not expect GCE models to produce the S1L in the O three-isotope plot. If we follow the common approach of renormalizing the GCE model so that the ratios predicted at solar metallicity are scaled to the terrestrial values, we find that the slope of the predicted GCE line is still different from unity, specifically, it is equal to 1.065 if we consider all the points along the GCE curve in Fig. 2 or to 1.155 if we consider only the five points with metallicity closest to solar.

In terms of observations of the GCE evolution of the $^{18}O/^{17}O$ ratio, the solar value of 5.2, which may represent the interstellar medium (ISM) 4.6 Gyr ago, can be compared to the ratios observed in MCs and YSOs and those measured in stardust oxide grains, which originated in stars that formed and evolved prior to the formation of the Solar System more than 4.57 Gyr ago. It has been argued that the true ISM $^{18}O/^{17}O$ ratio at the time of the formation of the Solar System was close to the value of ~ 4 observed in MCs and YSOs located at the same distance from the Galactic centre as the Sun and that the higher solar ratio could be the result of pollution by massive stars and/or SNII (e.g., Prantzos et al. 1996; Young et al. 2011). In this case a GCE line of slope ~1 would have to be shifted by +300‰ in $\delta^{17}O$ in the O three-isotope plot to pass through the observations of MCs and YSOs (Young et al. 2010). This is in contradiction with the GCE interpretation of the S1L, which assumes that GCE passes through the solar composition.

An alternative explanation is that the solar $^{18}O/^{17}O$ ratio represents the true ISM at the time of the formation of the Sun and the lower ratio observed in MCs and YSOs is the result of GCE in the past 4.6 Gyr evolving towards lower $^{18}O/^{17}O$ ratios as more $^{17}O$ is produced by the longer-living stellar objects (Gaidos, Krot & Huss 2009; Nittler & Gaidos 2012). This implies a slope different from unity on the O three-isotope plot. This second hypothesis is supported by the analysis of stardust oxide grains, which indicate that the solar $^{18}O/^{17}O$ ratio was typical for its age (Nittler 2009). Most of the O compositions of stardust grains are easily interpreted as the signature of the first dredge-up on the red giant branch, which can only increase $^{17}O/^{16}O$, while keeping $^{18}O/^{16}O$ almost unchanged (see, e.g., Fig. 1 of Nittler 2009). If the parent stars of the grains started with an initial $^{18}O/^{17}O$ ratio lower than solar, to explain the data we would need to invoke an unknown stellar nucleosynthetic process that decreases $^{17}O$ while keeping $^{18}O$ constant. Observations of MCs and YSOs located at a range of distances from the Galactic centre show that the $^{18}O/^{17}O$ ratio increases with the distance (Fig. 1 of Young et al. 2011). These spatial variations of the $^{18}O/^{17}O$ ratio may indicate that this ratio could be also affected by temporal variability.

We do not pursue here which of the two hypotheses proposed to explain the different $^{18}O/^{17}O$ ratio observed in MCs and YSOs and in the Solar System is correct, but we note that the GCE interpretation of the S1L is in disagreement with both of them. The hypothesis that states that the difference is due to the $^{18}O/^{17}O$ ratio evolving to different values directly contradicts the assumption that the $^{18}O/^{17}O$ ratio is constant required for the GCE model to explain the S1L. On the other hand, the hypothesis that states that the solar ratio does not reflect the true ISM $^{18}$O/$^{17}$O ratio at the time when the Sun formed, but is due to SNII pollution, is also in contradiction with the GCE interpretation of the S1L, as in this case GCE does not even pass through the solar O composition.

**3. The effect of incomplete mixing of stellar ejecta in the ISM**

When discussing small isotopic variations due to GCE the effect of incomplete mixing of stellar ejecta on the composition of the ISM needs to be considered. Lugaro et al. (1999) and Nittler (2005) have analyzed this effect in relation to the Si and Ti isotopic compositions of stardust silicon carbide (SiC) grains and the O composition of oxide grains. In these works a simple Monte Carlo model was used to represent random selection of stellar ejecta resulting in possible different compositions in different regions of the ISM. The spread obtained along the standard average GCE of the Si and Ti isotopes can be compared to the stardust data to set the free model parameters (Fig. 2 of Nittler 2005): the number of contributing supernovae to $N_{SN}$ = 70 and the dilution factor of the ejecta to $a$ = 5.5 x $10^{-6}$ $M_\odot^{-1}$, i.e., the ejecta from each star are diluted with an IMS mass of ~ 1.8 x $10^5$ $M_\odot$. The resulting variations in the Si and Ti isotopic compositions are ~ 5-10%. This is insignificant when compared to the overall GCE effect, but of great impact when considering the same-order spread in the Si and Ti isotopic compositions resulting from high-precision measurements of stardust grains (Lugaro et al. 1999, Nittler 2005) and, in our context, the 6% variations of O isotopic ratios in the Solar System.

We have used the same model of Lugaro et al. (1999), with the values of the free parameters reported above that match the stardust data, to analyze the impact of incomplete mixing of stellar ejecta on the GCE evolution of the O isotopes. We used SNII yields for stars in the mass range 11 < M/ M$_\odot$ < 40 from Woosley & Weaver (1995) and supernovae of Type Ia yields (SNIa, resulting from the thermonuclear explosion of C-O white dwarves) from Thielemann et al. (1986). We scaled $^{17}$O from SNII by 0.12 as suggested by Nittler (2005) to reproduce the models by Rauscher et al (2002), where $^{17}$O is lower due to updated rates of the $^{17}$O+p reactions. We also included the contribution of low- and intermediate-mass stars from Karakas (2010; 1.25 < M/ M$_\odot$ < 6.5) and of Super-AGB stars from Siess (2010; 9 < M/ M$_\odot$ < 10.5). With these choices for the yields our basic model reproduces the requirement that the overall GCE follows a S1L. We added to different choices of initial abundances the yields (diluted by *a*) from 70 supernovae (80% as SNII resulting from stars of mass > 11 M$_\odot$ and 20% as SNIa) and 1191 low- to intermediate-mass stars (M < 11 M$_\odot$, evaluated using the Salpeter initial mass function). The yields are drawn randomly with probabilities associated to each stellar mass according to the Salpeter initial mass function. We repeated this procedure 500 times to derive 500 possible compositions produced by incomplete mixing of random stellar ejecta. The results are presented in Fig. 3.

Adding random stellar ejecta to an initial composition at $\delta^{18}$O = $\delta^{17}$O = -85 (black hexagon on Fig. 3, chosen so that the resulting O compositions average to that of the Sun) we obtained a set of new compositions, which spread by roughly ±20‰ around an average at $\delta^{18}$O = $\delta^{17}$O ~ -60‰ (red open dots). We take these to represent the Solar

Nebula gas component. These different local compositions are expected to mix within the timescale by which material at a given Galactic radius is homogenized, of the order of 350 Myr (de Avillez & Mac Low 2002), and to produce the average O ratios that we have used as the initial composition for our next Monte Carlo calculations. We repeated this procedure until we reached compositions with an average $\delta^{18}O = \delta^{17}O \sim 0$ (blue points in Fig. 3). We take these to represent the Solar Nebula dust component. Overall, subsequent generations of heterogenous mixing result in a GCE evolutionary path where a spread is superimposed on to the homogeneous GCE path. That the parent stars of stardust SiC grains, in particular the Si versus Ti anomalies, have kept a record of ISM heterogeneities indicates that star formation occurs within a timescale shorter than the homogenization timescale and/or that new heterogeneities are created as old heterogeneities are erased. The stellar yields considered in our models could be shed in the ISM in the form of both gas and stardust, so we predict a spread due to heterogeneities in the GCE dust component in case this component comes from stardust, MC dust, or a combination of both. We stress that though our model is very basic, it is observationally supported by the Si and Ti compositions of stardust SiC grains.

The majority of the local ISM compositions produced by incomplete mixing are not located on the S1L. Of the set of 500 points we computed for each initial composition ~10% and ~27% satisfy the $\delta^{18}O = \delta^{17}O$ condition within 1 and 3 permil, respectively. In order to obtain the possible mixing combinations (of which one might be the S1L) between the dust and the gas components we need to connect one point drawn from the compositions representing the gas component (red points in Fig. 3 with average ~ -60‰) to one point draw from the compositions representing the gas component (blue points in Fig. 3 with average ~ 0). Out of all the possible lines we obtain a S1L only when both the connecting points have $\delta^{18}O = \delta^{17}O$. The probability of this to occur is 1% and 7%, for a line located within 1‰ and 3‰, respectively, of the line of slope exactly =1. The example shown in Fig. 3 is illustrative only, but clearly the same conclusion applies if the dust component is assumed to derive from dust of composition $\delta^{18}O$ and $\delta^{17}O \gg 0$. As time passes the degree of mixing increases and the level of heterogeneity in each of the components decreases. The final result will depend on the timescale of formation of the two different reservoirs. We address the issue of timescales in the next section.

We note that more sophisticated GCE models are currently being developed, which also include hydrodynamics and feedback effects (e.g., Kobayashi & Nakasato 2011, Pilkington et al. 2012). These more detailed models perform better at matching the large intrinsic spread in the age-metallicity relation observed in the solar neighbourhood (e.g., Holmberg et al. 2007). Future chemodynamical models of the O isotopic evolution in the Galaxy may predict an even larger spread of values at any given time than that obtained only considering incomplete mixing of stellar ejecta.

**4. The GCE timescale of the O isotopic evolution**

The main other requirement for the GCE scenario to explain the S1L is that the dust component of the Solar System is somewhat younger than the gas component. The timescale at which GCE progresses is relatively long, of the order of the age of the Galaxy. One would expect that to modify the O isotopic composition by only 6% at around the time of the formation of the Sun requires a considerable amount of GCE time. The order of magnitude of this timescale can be obtained by translating the required change in [Fe/H] to the elapsed time using, e.g., Fig. 12 and Fig. 7 of Timmes et al (1995) or Fig. 17 and Fig. 11b of Kobayashi et al. (2011). We derive a timescale of the order of 1 Gyr, which means that in order to explain the solar system systematics the dust component in the Solar System should have formed from stellar ejecta roughly 1 Gyr after the ejecta that contributed to the gas component. We note that this assumes that the GCE dust component is derived completely from pure stardust and MC dust. If the dust component originated from a combination of stardust, MC dust, and Solar-System condensates, then the stardust and MC dust are required to have $\delta^{18}O$ and $\delta^{17}O \gg 0$ to balance the composition of the Solar-System condensates, which also carry the signature of the gas component. This would imply a timescale longer than 1 Gyr.

Since the GCE timescale is longer than both the timescale required for the ISM to be well mixed (~$10^8$ yr, de Avillez & Mac Low 2002) and the timescale for dust formation in the ISM (~$10^6 - 10^7$ yr, Zhukovska et al. 2008) the expectation is that MC dust should have the same composition as the MC gas. It follows that for the GCE scenario to work it is required that the signature of more recent stellar ejecta is kept somehow separated from the ~ 1 Gyr older gas in MCs. The segregation has to hold until the Solar System forms when dust and gas can mix together to produce the S1L. It is not clear how this segregation could occur for MC dust.

A possible solution is that the dust component of the Sun was dominated by stardust delivered to the ISM by younger stellar ejecta (Meyer 2009). The composition of this stardust will have to carry the signature of GCE and average to $\delta^{17}O \sim \delta^{18}O \gg 0$. However, O is strongly affected by stellar nucleosynthesis and it follows that the average composition of stardust is expected to (1) carry a much stronger signature of stellar nucleosynthesis than of GCE and (2) be determined by the rate of dust production - also in terms of different sizes - in different stellar objects. Within the collection of stardust currently available, probably sampling the most common and largest size stardust, the vast majority of it shows the signature of an origin in AGB stars, which are also well known to be the predominant contributors to stardust in the Galaxy (Gehrz 1989). Accordingly, as mentioned in Sec. 2, the O compositions of most stardust show the order of magnitude variations with respect to the solar composition expected by the operation of nucleosynthesis in red giant and AGB stars and average to a composition enhanced in $^{17}O$ and depleted in $^{18}O$ (see, e.g., Fig. 1 of Nittler 2009). Only a very small fraction of stardust oxide grains (Group 4 in Nittler et al. 1997, 2008) show large $^{17}O$ and $^{18}O$ enrichments, and these are explained as the signature of an origin in core-collapse supernovae rather than the imprint of GCE. Finally, the GCE ~1 Gyr timescale is of the same order as the survival time of dust in the ISM (Jones et al. 1996, Gyngard et al. 2009) and one would expect stardust and MC dust present at the time of the formation of the Sun to have ages ~ 1 Gyr older than the gas.

**5. Discussion and conclusions**

We have analyzed in detail the basic assumptions of the GCE interpretation of the S1L against current GCE models and observations. The requirement that the $^{18}O/^{17}O$ ratio should evolve as a constant is found to be unlikely on the basis of the most recent GCE models and when considering the effect of incomplete mixing of stellar ejecta. Furthermore, the GCE scenario is in contradiction with both the current hypotheses to explain the difference between the $^{18}O/^{17}O$ ratio observed in the Solar System and in MCs and YSOs. The requirement that the dust present at the formation of the Solar System was younger than the gas is in contradiction (a) in the case of MC dust, with the basic timescales involved in determining its formation and composition - GCE, ISM mixing, and dust formation in the ISM - and (b) in the case of stardust, with the results of the processes that mostly affect its average composition: stellar nucleosynthesis and dust condensation in stellar outflows.

A strict correlation has been observed in carbonaceous chondrites (CC) between variations in $\varepsilon^{54}Cr$, i.e., the $^{54}Cr/^{52}Cr$ ratio with respect to the solar value per ten thousand, and in $\Delta^{17}O = \delta^{17}O - 0.52\,\delta^{18}O$, i.e., the distance from the TFL, unaffected by mass-dependent fractionation (Shukolyukov & Lugmair 2006, Trinquier et al. 2007, Yin et al. 2009). This correlation has been invoked as evidence for the GCE scenario because $^{16}O$-poor stardust is expected to be enriched in the partially secondary $^{54}Cr$ (Yin et al. 2009, Krot et al. 2010). Dauphas et al. (2010) and Qin et al. (2011) have identified < 0.2 μm-sized oxide stardust grains carrying huge anomalies in the $^{54}Cr/^{52}Cr$ ratios, up to >10 times the solar value, however, correlated O data are not available yet. Nevertheless, there are other possibilities for the origin of the $\varepsilon^{54}Cr$ versus $\Delta^{17}O$ correlation. For example, it could be ascribed it to the presence of the common AGB stardust. Enrichments in the CC matrix fractions both of Mo isotopes due to the *slow* neutron capture (*s*) process in AGB stars (Dauphas et al. 2002) and of $^{54}$Cr, which can also be produced by the *s* process in AGB stars ($^{54}$Cr/$^{52}$Cr ratios up to +20% higher than solar, Lugaro et al. 2004) argue for such common source for these isotopes. As discussed above, AGB stardust carries the signature of $^{17}$O enhancements. Another option is that these correlations may simply reflect the amount of CAI material in the bulk meteorites, with CAIs carrying the anomalies in O and Cr.

In conclusion, for the GCE model to be viable one would need to (1) find a third, plausible hypothesis consistent with the GCE scenario to explain the $^{18}$O/$^{17}$O observations presented in Sec. 2, and (2) find a way to keep the gas and the dust component in the Solar Nebula separated. This would require either that the GCE evolution of the O isotopes is much faster than currently predicted, so that the stellar ejecta from which the MC dust formed occurred less than 350 Myr after the gas component is established, or that there existed an abundant but as yet unidentified stardust component with an O composition such that its average is $\delta^{17}O \sim \delta^{18}O \gg 0$. Both possibilities seem unlikely at the present. Even assuming one of them is correct, there would still be the issue that incomplete mixing of stellar ejecta results mostly in compositions that do not lie on the S1L. Due to all these problems, we rule out GCE as a likely explanation for the S1L. More effort should be put into providing a working model based on PSS.

Acknowledgements

We thank Chiaki Kobayashi for discussion and data for Figure 2 and Martin Asplund for discussions on O stellar observations. We are grateful to the referee Larry Nittler for pointing out several improvements to the manuscript. ML acknowledges support from the Australian Research Council via the Future Fellowship and from Monash University via the Monash Fellowship. KL acknowledge funding and support from CSIRO Astrophysics and Space Sciences (CASS).REFERENCES

Asplund, M., Grevesse, N., Sauval, A. J., & Scott, P. 2009, ARA&A 47, 481

Clayton, D. D. 1988, ApJ, 334, 191

Clayton, D. D., & Pantelaki, I. 1986, ApJ, 307, 441

Clayton, R. N. 2002, Nature 415, 860

Clayton, R. N., Grossman, L. & Mayeda, T. K. 1973, Science 182, 485

Clayton, D. D., & Nittler, L. R. 2004, ARA&A, 42, 39

Clayton, R. N., Onuma, N., Grossman, L., & Mayeda, T. K. 1977, EPSL, 34, 209

Dauphas, N., Remusat, L., Chen, J. H., Roskosz, M., Papanastassiou, D. A., Stodolna, J., Guan, Y., Ma, C. & Eiler, J. M. 2010, ApJ, 720, 1577

Dauphas, N., Marty, B., & Reisberg, L. 2002, ApJ, 569, 139

de Avillez, M. A., & Mac Low, M.-M. 2002, ApJ, 581, 1047

Gaidos, E., Krot, A. N., & Huss, G. R. 2009, ApJL, 705, L16320

FIGURE CAPTIONS

Figure 1: A schematic picture representing a sample of O isotope compositions intrinsic to the Solar System. Compositions range from rare meteoritic inclusions that are $^{16}$O-rich ($\delta^{18}$O ≈ -80 ‰ for chondrule a006 from Acfer 214, Kobayashi et al. 2003) to the $^{16}$O-poor Insoluble Organic Material derived from the matrix of Yamato-793495 (Hashizume et al. 2011) with $\delta^{17}$O and $\delta^{18}$O up to +400 ‰ (though these results have not been replicated by other researchers, Larry Nittler, personal communication). The other materials represented are Cosmic Symplectite of predominantly Fe oxide from Acfer 094 (Sakamoto et al. 2007); Lunar Metal - surface oxygen in lunar metal grains (Ireland et al. 2006); CAIs and chondrules (Clayton et al. 1977), which also include the extreme compositions measured from Murchison hibonite inclusions (Ireland et al. 1992); Sun - inferred solar composition from solar wind measurement (McKeegan et al. 2011) and the TFL.

Figure 2: The GCE of the O isotopic ratios in the Solar Neighborhood as a function of [Fe/H] values ranging from -2.6 to +0.14 computed by Kobayashi et al. (2011) as compared to the S1L and the TFL.

Figure 3: Oxygen three-isotope plot showing the GCE of the O isotopic ratios resulting from incomplete mixing of stellar ejecta in the ISM. Random stellar ejecta added to an initial composition at $\delta^{18}$O = $\delta^{17}$O = -85 (black hexagon) result in the red open dots, taken to represent possible Solar Nebula dust compositions. Each point represents one of the 500 computed local ISM compositions. The large red hexagon on the S1L at ~ -60 is the average of these compositions. Random stellar ejecta added to this average result in the compositions represented by the cyan points. The same procedure is applied moving to the green and then the blue points, taken to represent possible Solar Nebula dust compositions, whose average is ~ 0. The black line with slope > 1 connecting a red and a blue point of relatively extreme compositions represents an example of the ~$10^5$ possible slopes generated by mixing a random gas-component point to a random dust-component point. In this exercise it took three steps to move from the gas to the dust composition. This is determined by the choice of the stellar yields, but it is consistent with the GCE timescale of ~1 Gyr discussed in Sec. 4, when taking three times the ISM mixing time of ~350 Myr.

Figure1

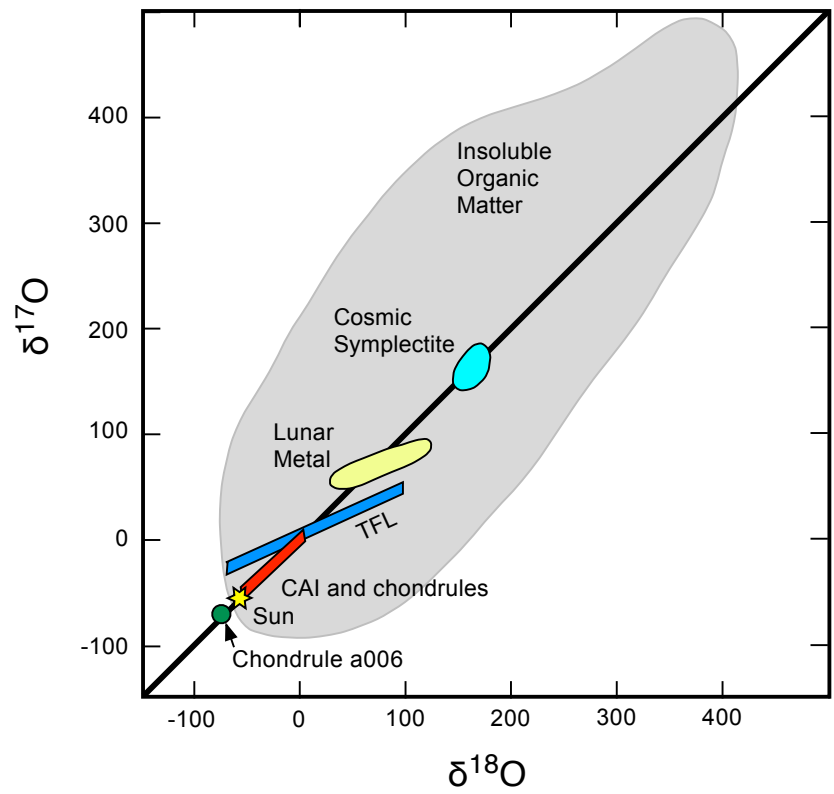

Figure 2

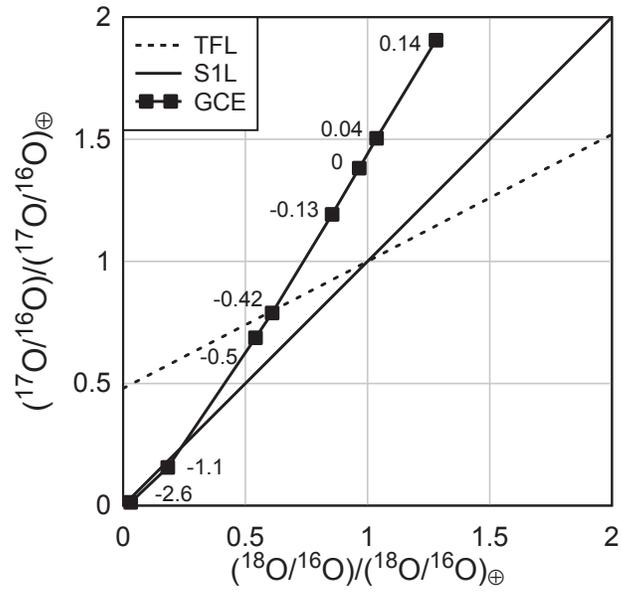

Figure 3

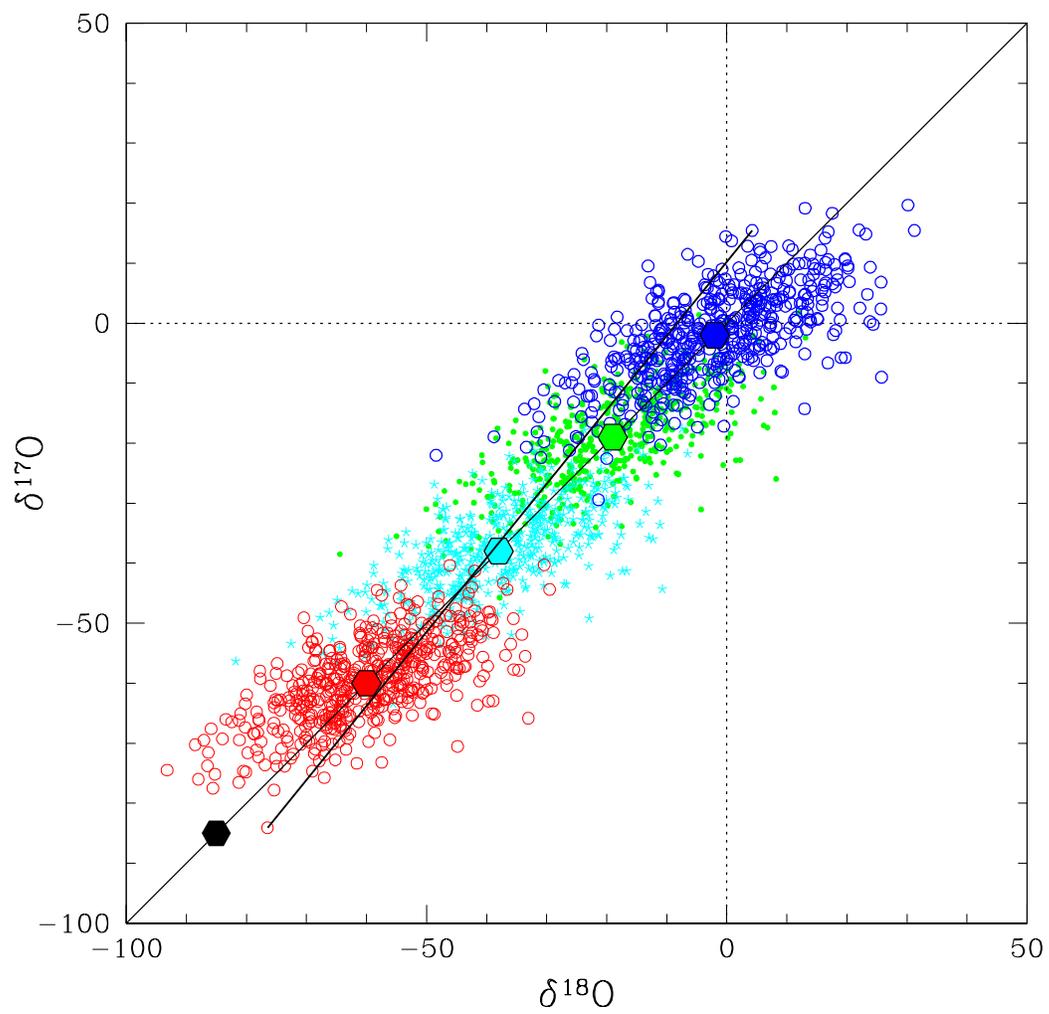